\documentclass[12pt]{article}
\usepackage{graphicx}

\def\pt{p_T}
\def\kt{k_T}
\def\dis{distribution}
\def\tj{2-jet recombination}
\def\oj{1-jet fragmentation}

\begin{document} 

\begin{center}  {\Large {\bf Effects of Near-side Two-jet Recombination on Away-side Correlation at LHC}}
\vskip .75cm
 {\bf Rudolph C. Hwa$^1$ and C.\ B.\ Yang$^{1,2}$}
\vskip.5cm
{$^1$Institute of Theoretical Science and Department of
Physics\\ University of Oregon, Eugene, OR 97403-5203, USA\\
\bigskip
$^2$Institute of Particle Physics, Hua-Zhong Normal
University, Wuhan 430079, P.\ R.\ China}
\end{center}

\vskip.5cm
\begin{abstract} 
The creation of partons with transverse momenta around 10 GeV/c in Pb-Pb collisions at LHC is expected to be so copious that the parton density may be high enough for neighboring jet cones to overlap. If the shower partons in adjacent jets can recombine to form trigger particles, the distribution of hadrons  on the away side should be distinguishable from that due to the recoil of a  single harder parton that can produce a similar trigger particle by fragmentation. The two-particle azimuthal correlation on the away side is generated by simulation of the two trigger types. Moments of the correlation distributions are calculated to exhibit their differences. A measure that can be determined by suitable analysis of the LHC data  is suggested. It may reveal the footprint of two-jet recombination.

\end{abstract}
\vskip0.5cm

\section{\large Introduction}
As the Large Hadron Collider (LHC) starts producing data on Pb-Pb collisions, many predictions based on the extrapolation of known observables at Relativistic Heavy-ion Collider (RHIC) will be tested. Measurable quantities such as multiplicity density and elliptic flow have already been reported \cite{alice}; they provide some realistic picture of the global properties at the new energy  and density frontier. There are other features of particle production at LHC that will take some time to be revealed because of the detailed data analysis required. They hold prospects of uncovering new physics that cannot occur at lower energy and therefore cannot be predicted quantitatively by simple extrapolations from RHIC. The subject of our investigation described in this article concerns a new dynamical process of that kind. It may be discovered at LHC, but has not been detected at RHIC.

The production of a high-$p_T$ particle at RHIC is a rare event. Given a hard parton at high $\kt$, the hadron momentum in the jet can usually be calculated by fragmentation in isolation, i.e., independent of other partons in the vicinity of the jet. That scenario must be modified if jets are not rare. At LHC partons with $\kt\sim 10$ GeV/c are expected to be produced copiously. When the local density of such partons is high enough, the shower partons generated by such closely packed initiating partons may well overlap spatially. If it occurs, then the hadronization of those overlapping shower partons may take place by recombination, generating particles at $\pt\sim 10$ GeV/c. Such particles, hereafter referred to as products of \tj, can become a strong component that contaminates the result one expects from 1-jet fragmentation arising from harder partons at higher $\kt$. In Ref.\ \cite{hy} we have carried out a preliminary study of some properties of \tj\ at LHC. Although there is no reliable estimates of the parton density and the probability of recombination of shower partons from adjacent jets, it is possible to calculate the $p/\pi$ ratio once an assumption is made on the overlap probability because coalescing quarks to form proton and pion at the same $\pt$ have definite momentum-ratios of the order of 1/3 to 1/2. It is found that the $p/\pi$ ratio can be as large as 20, if \tj\ is favorable \cite{hy}. That would be a unique signature that awaits particle identification.

In a separate investigation the process of \tj\ is found to increase significantly the nuclear modification factor $R_{AA}$, making it possible to exceed 1 \cite{hy1}. Moreover, a scaling behavior (consisting of separate variations of centrality and azimuthal angle $\phi$), found to be valid at RHIC, is shown to be badly broken by \tj\ at LHC.

While the results in \cite{hy,hy1} are striking, there is a need for a more direct test of the physics of \tj. Single-particle $\pt$ spectra of various hadronic types cannot by themselves determine the nature of hadronization, as evidenced by the lack of general consensus on that nature even at RHIC energies where precision data are abundant \cite{ph,st} and deviations from specific theoretical expectations are referred to as anomalies. Two-particle correlations probe deeper the underlying dynamical processes, but no observables examined so far at RHIC offer the feasibility of distinguishing different types of hadronization process at LHC. Our goal in this paper is to find a measure that may be useful in the analysis of LHC data in any attempt to find unusual signals, which should be of interest in their own right in general, but may provide more direct evidence for \tj\ in particular.

The idea is basically very simple. For a trigger with hadronic momentum $\pt\sim 10$ GeV/c, if it is due to the fragmentation of a hard parton, then the parton momentum $\kt$ should be much higher than $\pt$, and therefore higher than those of the two partons that recombine to form the trigger. If the latter case arises from two jets on the near side, their recoil partons that are directed toward the away side must each have lower momentum than that of the single parton in the former case. The patterns of away-side hadrons in the two cases should exhibit recognizable differences. If the threshold of away-side momenta is set appropriately high to suppress the medium response, single-jet recoil should have only one jet peak, while two-jet recoil should have two jet peaks. The measurable patterns for the two cases should reveal significantly different two-particle correlations. That is the problem that we shall focus on. We aim to construct an observable quantity that can convey the information on whether \tj\ plays an important role in the formation of the trigger on the near side, bearing in mind that an experimental measure cannot directly discriminate between different hadronization processes.

Since the construction of an experimental measure is a matter of relevance and detectability, and need not be based on the precision of quantitative demonstration, we shall perform simple simulation of the events that embody our physical ideas. We shall emphasize the fluctuation of azimuthal correlation of two particles on the away side and calculate the moments of the correlation \dis s of the two event sets corresponding to \oj\ and \tj\ for the trigger. From the result of the simulation we suggest a quantity that can be determined by analyzing experimental data; the behavior of that quantity as function of trigger momentum may reveal the presence or absence of \tj.

\section{\large Triggers in 1-jet fragmentation and 2-jet recombination}
From our previous formulation of the problem of single-particle production at high $p_T$ due to 1-jet fragmentation and 2-jet recombination at LHC \cite{hy, hy1}, we now treat the high-$\pt$ particle as a trigger and consider the effect of the recoil parton(s) in generating  two-particle correlation on the away side. In simulating events of interest we shall put in only the essence of trigger formation based on the two different types of hadronization process.  Since trigger bias favors the hard-scattering point to be close to the near-side surface \cite{hy2}, we assume that the point is within a layer of the medium near the surface, and set its thickness to be 0.2 of the overall path length $\xi$ of the medium, i.e., the point of hard-parton creation is at $\zeta$ from the near-side surface with $0<\zeta < 0.2 \xi$.  Here, $\xi$ is the dimensionless dynamical path length found in \cite{hy1} in terms of which the nuclear modification factor $R_{AA}(\phi, b)$ exhibits a remarkable scaling behavior at RHIC.
Numerically, our definition of $\xi$ used here is twice that defined in \cite{hy1} where the length is measured from the center of the nuclear overlap. As the parton moves from the creation point to the surface, it loses energy with 
the momentum-degradation factor that can be written as $\exp(-\zeta)$.  We assume that similar energy-loss consideration can be used at LHC.  Thus the hard-parton distribution at the surface with momentum $q$ in the transverse plane at mid-rapidity is
\begin{eqnarray}
F_i(q) = \int dkkf_i(k)  G(k,q,\zeta),     \label{1}
\end{eqnarray}
where $f_i(k)$ is the momentum distribution at the point of creation and $G(k,q,\zeta)$ is the degradation factor
\begin{eqnarray}
G(k,q,\zeta) = q\delta (q-ke^{-\zeta}).     \label{2}
\end{eqnarray}
For parton of type $i$ the distribution $f_i(k)$ at LHC is parametrized as \cite{sg}
\begin{eqnarray}
f_i(k) = f_0(1+k/B)^{-\beta}.        \label{3}
\end{eqnarray}
For simplicity we assume that the hard parton is directed normal to the surface so that the azimuthal angle $\phi$ can be ignored.  The complications associated with $\phi$ and centrality have been fully investigated in \cite{hy1}.

The 1-jet and 2-jet contributions to a trigger particle with momentum $p_t$ are, respectively, given by
\begin{eqnarray}
{dN^{1j}\over p_tdp_t} = \int {dq\over q} \sum_i F_i(q)H_i(q,p_t),     \label{4}
\end{eqnarray}
\begin{eqnarray}
{dN^{2j}\over p_tdp_t} = \int {dq\over q} {dq'\over q'} \sum_{i,i'}F_i(q)F_{i'}(q')H_{ii'}(q,q',p_t),    \label{5}
\end{eqnarray}
where $H_i$ and $H_{ii'}$ are two distinctly different hadronization functions.  In Eq.\ (\ref{5}) there are two hard partons ($i$ and $i'$) created, the probability for which is expressed by $F_i(q)F_{i'}(q')$ that is a product of independent probabilities of producing partons $q$ and $q'$ at the surface.  The creation points may be different, each traveling independent $\zeta$ and $\zeta'$ to get to the surface.  The directions of their trajectories can also, in principle, be different, but in order for them to contribute simultaneously to the formation of a single trigger particle with momentum $p_t$, it is necessary for them to be collinear.  The probability for that to occur is summarized by a parameter $\Gamma$ in the hadronization function $H_{ii'}(q,q',p_t)$ \cite{hy,hy1}.

In the 1-jet case $H_i(q,p_t)$ is just the fragmentation function, which has been treated as the recombination of two shower partons in the same jet \cite{hy3}
\begin{eqnarray}
H_i(q,p_t) = {1\over p_t^2} \sum_{jj'} \int {dq_1\over q_1} {dq_2\over q_2} S_i^j\left({q_1\over q}\right) S_i^{j'}\left({q_2\over q-q_1}\right) R_\pi(q_1,q_2,p_t),     \label{6}
\end{eqnarray}
where $S_i^j$ is the shower-parton distribution, derived from the fragmentation function $D(z)$ in  \cite{hy4}; $R_\pi$ is the recombination function for pion.  A parametrization of $S_i^j(z)$ is also given in \cite{hy4} for easy application.  The point of reiterating the details of $H_i(q,p_t)$ in Eq.\ (\ref{6}) is to show its difference from the 2-jet hadronization function
\begin{eqnarray}
H_{ii'}(q,q',p_t) = {1\over p_t^2} \sum_{jj'} \int {dq_1\over q_1} {dq_2\over q_2} S_i^j\left({q_1\over q}\right)S_{i'}^{j'}\left({q_2\over q'}\right) R_\pi^\Gamma(q_1,q_2,p_t),     \label{7}
\end{eqnarray}
where the two shower partons $j$ and $j'$ arise independently from two separate hard partons $i$ and $i'$ that are nearby, and they recombine to form a pion through $R_\pi^\Gamma(q_1,q_2,p_t)$ that depends on the overlap of the two jet cones.  The very small difference apparent in Eqs.\ (\ref{6}) and (\ref{7}) makes a huge difference in the probability of forming the pion, since $q_1$ and $q_2$ in Eq.\ (\ref{6}) are in the same jet of momentum $q$, which must be much larger than $p_t$, whereas $q_1$ and $q_2$ in Eq.\ (\ref{7}) are in separate jets of momenta $q$ and $q'$, neither of which is required to be greater than $p_t$.  At lower $q$ and $q'$ the hard-parton distributions $F_i(q)$ and $F_{i'}(q')$ in Eq.\ (\ref{5}) are higher than the one in Eq.\ (\ref{4}).  The consequence on the production of a high-$p_T$ trigger is significantly different as described in \cite{hy,hy1}, but our concern here is the consequence on the away-side particles whose azimuthal dependences should be measurably different for the two cases.

Our knowledge about $R_\pi^\Gamma(q_1,q_2,p_t)$ is insufficient to describe it in terms of the intrinsic and relative properties of the two jets.  A rough approximation is to write it as
\begin{eqnarray}
R_\pi^\Gamma (q_1,q_2,p_t) = \Gamma R_\pi(q_1,q_2,p_t),     \label{8}
\end{eqnarray}
where $\Gamma$ serves as the single parameter to represent the probability that the shower partons of two adjacent jets can recombine.  It should depend on the size of a jet cone and the density of jets, both of which depend on the jet momenta.  In our simulation of triggered events we need not specify $\Gamma$.  We consider all events where a particle $p_t$ is above a threshold, which we set to be 10 GeV/c.  We divide all those events into two sets:  1-jet and 2-jet.  We do not compare the probability of occurrences of the two cases.  Given a trigger in either set, we study the characteristics of the angular correlation of two particles on the away side.  From the difference between those characteristics, we aim to find a way to identify the relative abundance of the two cases in the experimental data, for which the value of $\Gamma$ cannot be controlled.

To simplify our simulation we consider only gluon jets, i.e., setting $i$ and $i'$ in Eqs.\ (\ref{4}) and (\ref{5}) to be $g$.  For every $k$ generated according to the distribution in Eq.\ (\ref{3}), we calculate $q$ according to Eq.\ (\ref{2}) for the gluon momentum at the surface.  We generate shower partons by use of $S_g^j(z)$ where $j$ is a light quark or antiquark, until the sum of all momentum fractions saturates 1.  Any pair in that set of shower partons can form a pion according to Eq.\ (\ref{4}).  If any produced pion has $p_t > 10$ GeV/c, a trigger is generated and we put the initiating parton of momentum $k$ at creation point $\zeta$ in the set of 1-jet triggered events.  For 2-jet triggered events we do the same but starting with two hard partons $k_1$ and $k_2$.  We demand that the shower parton in one jet recombines with a shower parton in the other jet according to Eq.\ (\ref{7}).  If the resulting pion has momentum greater than 10 GeV/c, we accept the event as one of 2-jet trigger and record $k_1$ at $\zeta_1$ and $k_2$ at $\zeta_2$.  From the accepted events of each type we use those initial states to generate configurations on the away side.

Since $f_g(k)$ is damped at high $k$ at an inverse power of $\beta = 5.6$ \cite{sg} and $S_g^j(z)$ is also suppressed at high $z$, as
\begin{eqnarray}
S_g^j(z) = Az^a(1-z)^b(1+cz^d),     \label{9}
\end{eqnarray}
where the parameters are tabulated in \cite{hy4} with $b=2.5$, the $k$ distribution is constrained on both the high and low sides.  For 1-jet events, $k$ is necessarily larger than $p_t$ (and hence $> 10$ GeV/c) due to momentum conservation, but for 2-jet events the sum $k_1+k_2$ must exceed $p_t$, not separately.  The recoil partons go through the bulk of the medium and experience different patterns of energy loss and angular deflection.

\section{\large Two-particle azimuthal correlation on the away side}
The observable we now consider is really three-particle correlation, one particle being the trigger, and the other two being on the away side.  For the events selected for trigger momentum $p_t > 10$ GeV/c we study the azimuthal correlation of the two particles on the opposite side.  We have argued in the preceding section that those triggered events can be of two types:  1-jet fragmentation and 2-jet recombination.  While it is possible for us to simulate those two types of events, an experiment cannot distinguish the two types of triggers.  It is our hope that there is enough difference in the observables on the away side to reveal a suitable criterion to disentangle the data and thereby identify the presence of 2-jet recombination on the near side.

The hard parton that recoils against a trigger parton traverses the major portion of the dense medium and can lead to various azimuthal patterns on the away side.  Two experiments at RHIC have given detailed analyses of the phenomenon \cite{ja, aa, ma}.  The main characteristics are that there is a peak at the $\Delta\phi = \pi$ region (called head or core) and double hump at $|\Delta\phi - \pi| \simeq 1.2$ region (called shoulder or cone), depending on the trigger $p_T^{\rm trig}$ and the associated $p_T^{\rm assoc}$.  Various theoretical models have been proposed to explain the structure \cite{iv}-\cite{kmw}, among which the mechanism of Mach-cone shock wave has been notable for the two peaks in the cone region \cite{hs, cst}.  There is no doubt that similar features will be seen at LHC.   In order to distinguish the signals that we seek, which are associated with the recoil partons themselves instead of the medium response to those hard partons (as is the case with Mach cone), we set the lower bound of $p_T^{\rm assoc}$ on the away side to be 5 GeV/c.  That should cut out most of the features of the thermal medium (which would correspond to the $TT$ component in the classification of the recombination model or to any conical behavior in a hydrodynamical model), and exhibit mainly the hadronization of the emerging partons ($TS + SS$ components).

Our simulation procedure for the away-side structure in $\phi$ distribution is simple and schematic, since there is a great deal of unknown about the medium created at LHC and our objective is only the search for some qualitative properties that can be suggestive of some measure useful for the analysis of Pb-Pb data.  The physical idea is very straightforward, which we describe first.  The procedure that follows to generate events is an approximate way to quantify the idea.

Consider first the case of 1-jet trigger.  Let a recoil hard parton of momentum $k$ be created at $\zeta$ from the near-side surface and directed at the opposite side.  It undergoes multiple scatterings, at each step of which it loses some momentum and is redirected at a random angle $\delta\phi$ relative to the original direction of that step, in a manner similar to the one described in  detail in a Markovian-Parton-Scattering Model \cite{ch}.
  If the angle $\delta\phi$ is given a Gaussian distribution, then upon emergence on the other side the total deflection $\Delta\phi$ (measured from $\pi$) is still a Gaussian with a wider width proportional to the square root of the number of steps taken.  The momentum $q$ of the parton at the surface is reduced from $k$ depending on the path length $\xi$. The exit parton creates a jet.  The momenta of hadrons ($p_b$) in the jet may be larger or smaller than $q$ depending on the hadronization process ($TS$ or $SS$ reco).  To avoid confusion, we use the same notation as in Ref.\ \cite{hy2} where $p_a$ is reserved for the transverse momentum of an associated particle on the near side, and $p_b$ for the away side.  If no particle in the jet has $p_b$ larger than a lower bound, which we set to be 5 GeV/c, then the event is rejected so as to avoid contamination by the hadronization of thermal partons.  For two-particle correlation on the away side, we require two hadrons both with $p_b > 5$ GeV/c.  The relative azimuthal angle between those two particles is restricted mostly to within the jet cone.  The jet direction may fluctuate over a wide range from event to event, but the dihadron relative angle within the same event fluctuates over a much narrower range.

For events with 2-jet trigger we have two hard partons $k_1$ and $k_2$ starting at $\zeta_1$ and $\zeta_2$.  For each parton the development of the trajectory toward the away-side surface is just as in the 1-jet case.  The two go through the medium independently, each exitting at an azimuthal angle with a Gaussian distribution that is uncorrelated to the other.  They hadronize separately, where we assume that the shower partons in the two jets do not recombine even if they overlap because of incoherence.  Thus not only the inter-jet angle fluctuates from event to event, but also the two hadrons in each event can fluctuate over a wide range of $\phi$ since they may originate from the two independent jets.  If they originate from the same jet, a possibility that is not ruled out if either $k_1$ or $k_2$ is large enough, then the relative $\phi$ is narrower as in the 1-jet case.

It is clear from the above description that the relative angle $\phi_1 - \phi_2$ between the two away-side hadrons is the important measure that contains the difference in the characteristics of the 1-jet and 2-jet event sets.  Denoting that relative angle by $\phi_-$, we should expect the distribution in $\phi_-$ to be wider in the 2-jet case than that in the 1-jet case.  Described below is a model generator that puts the measure on a quantitative basis.

Let us assume that at each step of a multiple-scattering Markovian process \cite{ch}, the $\phi$ deflection has a Gaussian distribution
\begin{eqnarray}
{\cal G}(\delta\phi) \propto \exp[-\delta\phi^2/2\sigma(\ell)^2]    \label{10}
\end{eqnarray}
where the width at the $\ell$th step is inversely proportional to the parton momentum $k(\ell)$ at that step, i.e., 
\begin{eqnarray}
\sigma(\ell) = k_0/k(\ell),     \label{11}
\end{eqnarray}
where $k_0$ is a constant which is to be set at 9 GeV/c, following the parametrization used in Ref.\ \cite{ch}.  The total net deflection has width-squared
\begin{eqnarray}
\sigma_i^2=k_0^2 \sum_{\ell} k(\ell)^{-2} = k_0^2 \int_0^{\ell_{\rm max}} {d\ell\over k(\ell)^2} = \left({k_0\over k_i}\right)^2 \left[e^{2(\xi-\zeta_i)}-1\right],     \label{12}
\end{eqnarray}
where the last equality is obtained by assuming momentum degradation to have the form \cite{hy1}
\begin{eqnarray}
k(\ell) = k_ie^{-\gamma\ell}     \label{13}
\end{eqnarray}
with $\ell$ being treated as the path length and $\xi = \gamma\ell$.  The subscript $i$ refers to the $i$th parton ($i=1, 2$) in the case of 2-jet events.  The parton momentum at the away-side surface is
\begin{eqnarray}
q_i = k_ie^{-(\xi-\zeta_i)}.     \label{14}
\end{eqnarray}
Each parton produces a jet of hadrons separately.

Since we consider only hadrons on the away side with $p_b > 5$ GeV, we can ignore $TT$ recombination and calculate only $TS + SS$ components.  For the $TS$ component that represent the medium effect on the shower partons, we have, as in the past \cite{hy1, hy3},
\begin{eqnarray}
{dN_{\pi}^{\rm TS}\over p_b dp_b}&=&{1\over p_b^2}  \int {dq_i\over q_i} F_g^{\rm away}(q_i)\widehat{\sf TS}  (q_i,p_b) ,  \label{15}  \\
\widehat{\sf TS}  (q_i,p_b)&=& \int dq'_1 Ce^{-q'_1/T} \int {dq'_2\over q'_2} S_g^j(q'_2/q_i) R_\pi(q'_1,q'_2,p_b),   \label{16}
\end{eqnarray}
where $F_g^{\rm away}(q_i)$ is the distribution obtained by simulation that results in Eq.\ (\ref{14}).  For LHC we set
\begin{eqnarray}
T = 0.5\ \rm GeV .    \label{17}
\end{eqnarray}
For the $SS$ component we use Eqs.\ (\ref{4}) and (\ref{6}), with $F_i(q)$ replaced by $F_g^{\rm away}(q_i)$, and $q_1(q_2)$ in (\ref{6}) treated as dummy variables, not to be confused with $q_i$ in Eq.\ (\ref{14}).

For each hard parton at $q_i$ we generate a set of shower partons until the sum of the shower parton momenta saturates $q_i$.  We then let those shower partons hadronize in accordance to the recombination procedure described above.  We let the pions generated to have an azimuthal distribution around the jet axis with a Gaussian width of 0.5 rad.  Only pions with $p_b > 5$ GeV/c are retained in the set for each jet.  For two-particle correlation we consider only the angles $\phi_1$ and $\phi_2$ of two pions, which may be in the two separate jets, or may be in the same jet.  That is, among all the particles in 1-jet or 2-jet sets, we choose randomly pairs of particles and study the density distributions $D(\phi_1,\phi_2)$.  
The maximum  dynamical path length in central collisions in Au-Au collisions  is  $\xi_{\rm max} = 2.9$, which is twice the maximum value ($0.6\times 2.4$) found in Ref.\ \cite{hy1}. It is in agreement with the value ($\beta L$) used in Ref.\ \cite{hy2} in a different approach to the problem. In our simulation for Pb-Pb collisions at LHC we consider only $\xi=3$. Figure 1 shows the 3D plots of the density  distributions of hadron pairs after simulating  $10^7$ events, most of which are rejected by our acceptance criteria. Among those accepted, most give rise to (a) 1-jet, and a smaller fraction to (b) 2-jet triggered events.  It is evident from (a) that the values of $\phi_1$ and $\phi_2$ in the 1-jet events are mainly along the diagonal with a Gaussian \dis\ centered at $\phi_1\sim\phi_2\sim\pi$. The few pairs in the off-diagonal corners arise from the artifact of identifying $\phi_i=0$ to $2\pi$; they belong to the diagonal beyond $(0,0)$ or $(2\pi,2\pi)$. 
In Fig.\ 1 (b) we see that the pairs are far more dispersed since the two particles may arise from separate jets. Nevertheless, there is a peak near $\phi_1\sim\phi_2\sim\pi$ because the two jets are separated by an angle that is  a Gaussian distribution centered at $\pi$ and the particles within each jet is also Gaussian distribute with width 0.5.
The difference between Fig.\ 1 (a) and (b) can also be seen clearly in 
their projections onto $(\phi_1,\phi_2)$ plane, as  shown in   Fig.\ 2. The dispersion from the diagonal in the  1-jet case  is evidently less than in the 2-jet case. To exhibit their differences more quantitatively, further analysis needs to be done on $D(\phi_1,\phi_2)$.

\section{\large Moments of $\phi_+$ and $\phi_-$}
Let us restrict our attention to the regions $\pi/2 < \phi_{1,2} < 3\pi/2$ relative to the trigger and define
\begin{eqnarray}
\phi_+ = \phi_1 + \phi_2 - 2\pi, \qquad \phi_- = \phi_1 - \phi_2,     \label{18}
\end{eqnarray}
so that both $\phi_+$ and $\phi_-$ are in the range ($-\pi, +\pi$).  After $D(\phi_1, \phi_2)$ is transformed to $D(\phi_+, \phi_-)$, let the probability of finding a pair of particles at $\phi_+$ and $\phi_-$ be
\begin{eqnarray}
P(\phi_+, \phi_-) = {D(\phi_+, \phi_-)\over \int_{-\pi}^\pi d\phi_+ \int_{-\pi}^\pi d\phi_- D(\phi_+, \phi_-)}     \label{19}
\end{eqnarray}
and its projection onto the $\phi_+$ and $\phi_-$ axes be
\begin{eqnarray}
P(\phi_\pm) = \int_{-\pi}^\pi d\phi_\mp P(\phi_+, \phi_-).     \label{20}
\end{eqnarray}
We show in Fig.\ 3 (a) $P(\phi_+)$, and (b) $P(\phi_-)$,  dashed lines for 1-jet trigger and solid lines for 2-jet trigger.  We see that there is no significant difference between the two trigger cases in $P(\phi_+)$, but 2-jet is noticeably wider than 1-jet in $P(\phi_-)$.  The reason is clear from the discussion in the preceding section that in the 2-jet case the particle pair can be from the two independent jets so their $|\phi_-|$ is larger than in the 1-jet case where the two particles are from the same jet.  Thus $P^{2j}(\phi_-)$ is wider than $P^{1j}(\phi_-)$, the latter having a width that is essentially the input width of the jet cone $(=0.5)$.  The width of $P^{1j}(\phi_+)$ and $P^{2j}(\phi_+)$ is largely determined by the net deflection angle of either jet, the former being slightly broader because the latter has a small enhancement near $\phi_+ \sim 0$ due to events with $\phi_1$ and $\phi_2$ on opposite sides of $\pi$.  Those events with large $|\phi_-|$ and small $|\phi_+|$ populate the off-diagonal corners of the $(\phi_1,\phi_2)$ plot in Fig.\ 2 (b), and are in the region where the double-hump structure has been seen in the RHIC data \cite{abe}.  However, since the threshold for $p_1$ and $p_2$ is 5 GeV/c, we do not anticipate the medium-response structure, such as Mach cone, to be present in the event sets chosen for LHC, and we see no enhancement in the off-diagonal corners in Fig.\ 2 (b).  
In the real experiment the lower bound for $p_1$ and $p_2$ can, of course, be varied to test that phenomenon.  What cannot be controlled in the data analysis is the degree of admixture of the 1-jet and 2-jet components, although it is expected that 1-jet trigger should dominate as $p_t$ is increased.

Our objective is to learn from the azimuthal properties on the away side the essential difference between the 1-jet and 2-jet cases so as to determine from the real data the probability of occurrence of 2-jet triggers.  To that end we consider the moments of $\phi_\pm$, since they put more weight on the wings of $P(\phi_\pm)$ where the difference is more pronounced.  Let us define
{\begin{eqnarray}
M_n(\phi_+)&=& \int d\phi_-\phi_-^n P(\phi_+,\phi_-),\\     \label{21} 
N_n(\phi_-) &=& \int d\phi_+\phi_+^n P(\phi_+,\phi_-).     \label{22}
\end{eqnarray}
There are significant amounts of fluctuations in our \dis s, as can be seen in Figs.\ 1-3, because in our simulation only about $5\times 10^4$ points have been recorded from $10^7$ events. However, the shapes of 
all the \dis s may be regarded as Gaussian. Thus, in calculating moments in Eqs.\ (21) and (\ref{22}) we use the results from fitting $P(\phi_+,\phi_-)$ by two-dimensional Gaussian \dis s.
We show in Fig.\ 4 (a) and (b) our simulated results for $M_n(\phi_+), n=2,4$, and in Fig.\ 5 (a) and (b) the results for $N_n(\phi_-), n=2,4$.  The solid lines are for 2-jet and dashed lines for 1-jet.  It is evident that the most drastic difference between the two event sets is the value of $M_n(\phi_+)$ at $\phi_+ = 0$, and $n=4$, where the 2j moment $M_4^{2j}(0)$ is more than 6 times larger than $M_4^{1j}(0)$.  The reason is the presence of more pairs off the diagonal in Fig.\ 2(b) than those in Fig.\ 2(a). 

To put the visual observation of the properties of Figs.\ 4 and 5 on a more quantitative basis, let us define the ratio of the off-diagonal to diagonal moments
\begin{eqnarray}
R_n^{(1j,2j)} = {M_n^{(1j,2j)} (0)\over N_n^{(1j,2j)}(0)} .     \label{23}
\end{eqnarray}
The values of the moments on the RHS can be determined by averaging $\phi_\pm$ over a narrow range around 0 to increase statistics.  We show in Fig.\ 6 the values of $R_n^{(1j,2j)}$ from our simulation.  For both $n=2$ and 4, the increase from 1-jet to 2-jet cases is obviously huge.  However, in real experiments 1-jet and 2-jet components cannot be put into separate event classes.  Consider then a mixture of the two components.  If $P(\phi_+,\phi_-)$ is written as a linear combination of the two trigger types
\begin{eqnarray}
P(\phi_+,\phi_-,\alpha) = P^{1j}(\phi_+,\phi_-) + \alpha P^{2j}(\phi_+,\phi_-),     \label{24}
\end{eqnarray}
where $\alpha$ is an unknown mixing parameter, then we have
\begin{eqnarray}
M_n(\phi_+,\alpha) &=& M_n^{1j}(\phi_+) + \alpha M_n^{2j}(\phi_+),\\     \label{25}
N_n(\phi_-,\alpha) &=& N_n^{1j}(\phi_-) + \alpha N_n^{2j}(\phi_-),     \label{26}
\end{eqnarray}
in terms of which we define
\begin{eqnarray}
R_n(\alpha) = {M_n(0,\alpha)\over N_n(0,\alpha)} .    \label{27}
\end{eqnarray}
In Fig.\ 7 we show from our simulation the dependence of $R_n(\alpha)$ on $\alpha$.  For $n=4$, $R_n(\alpha)$ increases by a factor larger than 4, when the 2-jet trigger becomes important.

Although $\alpha$ is not known in an experiment, the ratio of moments $R_n$ using the measured $P(\phi_+,\phi_-)$ defined in Eq.\  (\ref{19}) is a quantifiable observable in an experiment. We suggest that its value be measured and compared to the model curves in Fig.\ 7.  The precision of those curves is secondary compared to the fact that they vary over a wide range, so that the limits on the two ends are most likely to bracket the measured value of $R_n$.  We further suggest that the trigger momentum $p_t$ be varied over the range 8 to 16 GeV/c in 2 GeV/c bins.  Since 1-jet fragmentation becomes more important at higher $p_t$, we expect $p_t$ to play a role inversely proportional to the parameter $\alpha$.  If a plot of $R_n(p_t)$ vs $p_t^{-1}$ reveals some behavior similar to Fig.\ 7, then the experiment will have found some evidence in support of the possibility of 2-jet recombination. On the other hand, if $R_n(p_t)$ is found to be relatively independent of $p_t$, then the result can be used to rule out the presence of the effect of \tj\ in the away-side correlation. Such a possibility seems to be well worth the effort of analyzing the Pb-Pb data at LHC along the line suggested.

\section{\large Conclusion}

We have considered the problem of finding suitable experimental measure that can reveal the presence of \tj\ at LHC. We have been led to the study of two-particle correlation on the away side in order to take advantage of the nuclear medium in its role to enhance the difference between the two cases of 1-parton and 2-parton recoil. Associated particles on the near side would not have that advantage since they would have the same ambiguity that the trigger particle itself has in distinguishing the two hadronization processes. On the near side the two recombining jets must be nearly parallel and close-by, but the recoiling partons need not be so at the exit points of the away side.

Our simulated events are designed to display the properties of the medium effects on the two types of recoil partons. The accuracy of our description of the physics involved is unknown in the absence of any empirical information on the away-side phenomenon in Pb-Pb collisions at LHC. However, qualitative features of what we have found by quantifying the simulated results give light to the possibility of some useful measures that should be considered anyway in the analysis of real data from LHC, regardless of the motivation to search for evidences of \tj.

It is our suggestion that moments of the pair \dis\ in $(\phi_+,\phi_-)$, opposite the trigger, can reveal effectively the correlation structure characteristic of 1-parton or 2-parton recoil. Furthermore, the ratio of the off-diagonal moments to the on-diagonal moments, $R_n$, provides a succinct measure that can be determined experimentally on the one hand, and is shown in model calculation to have sensitive dependence on the nature of the hadronization process of the trigger on the other. In our simulation the lower bound of hadron momenta on the away side is set high to suppress the effects of low-$\pt$ medium response. The method can, however, be used also at any $\pt$ range if the focus is directed at precisely those effect that are suppressed in this study. In fact, the method should be applied to the analysis of RHIC data to quantify the double-hump structure on the away side already observed.

\section*{Acknowledgment}

This work was supported  in
part,  by the U.\ S.\ Department of Energy under Grant No. DE-FG02-96ER40972 and by the National Natural Science Foundation of China under Grant No.\ 10775057 and 11075061 and by the  Program of Introducing Talents of Discipline to Universities under No.\ B08033.

\begin{figure}[tbph]
\vspace*{-3cm}
\includegraphics[width=1\textwidth]{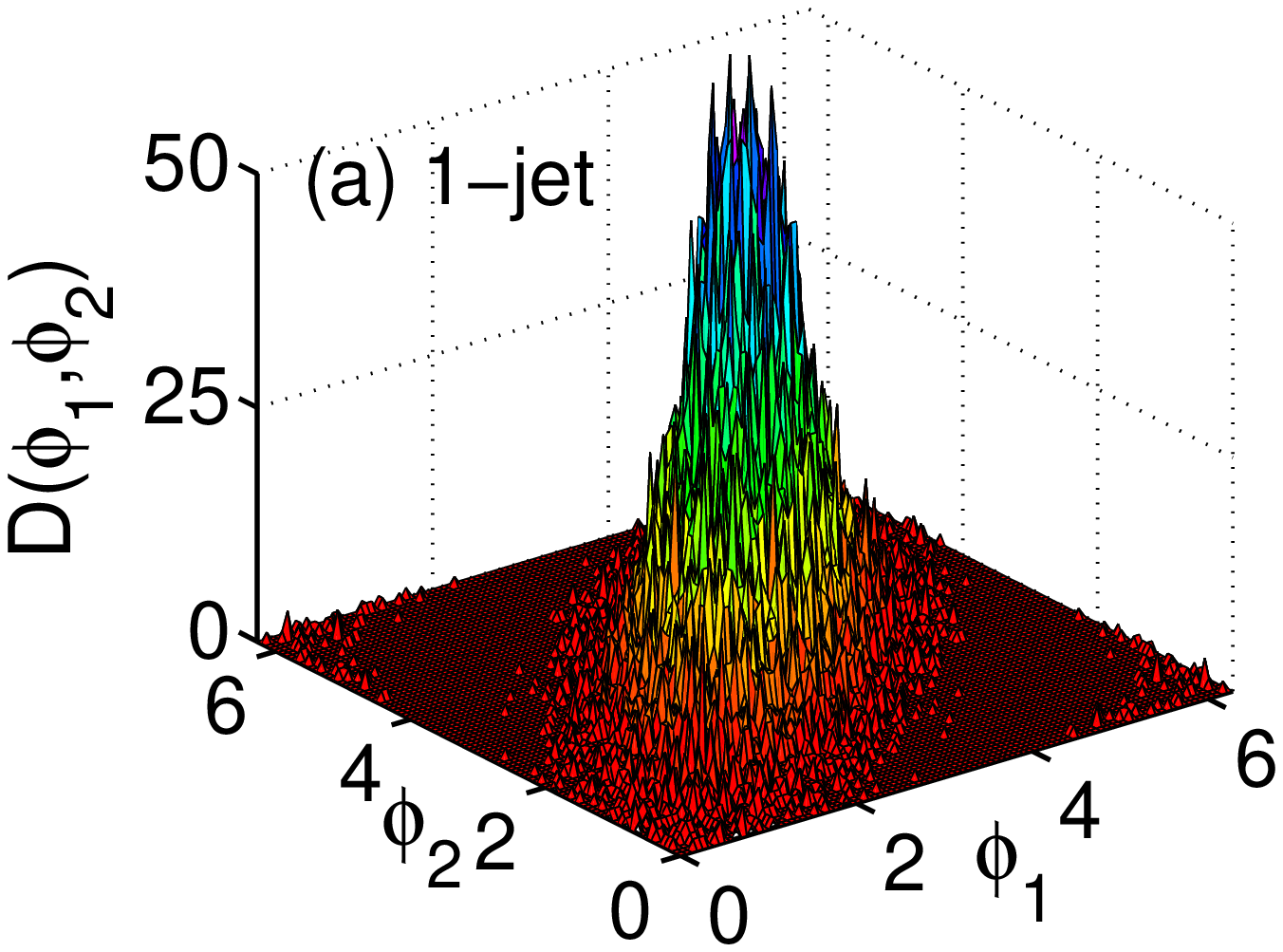}
\vspace{-1cm}
\includegraphics[width=1\textwidth]{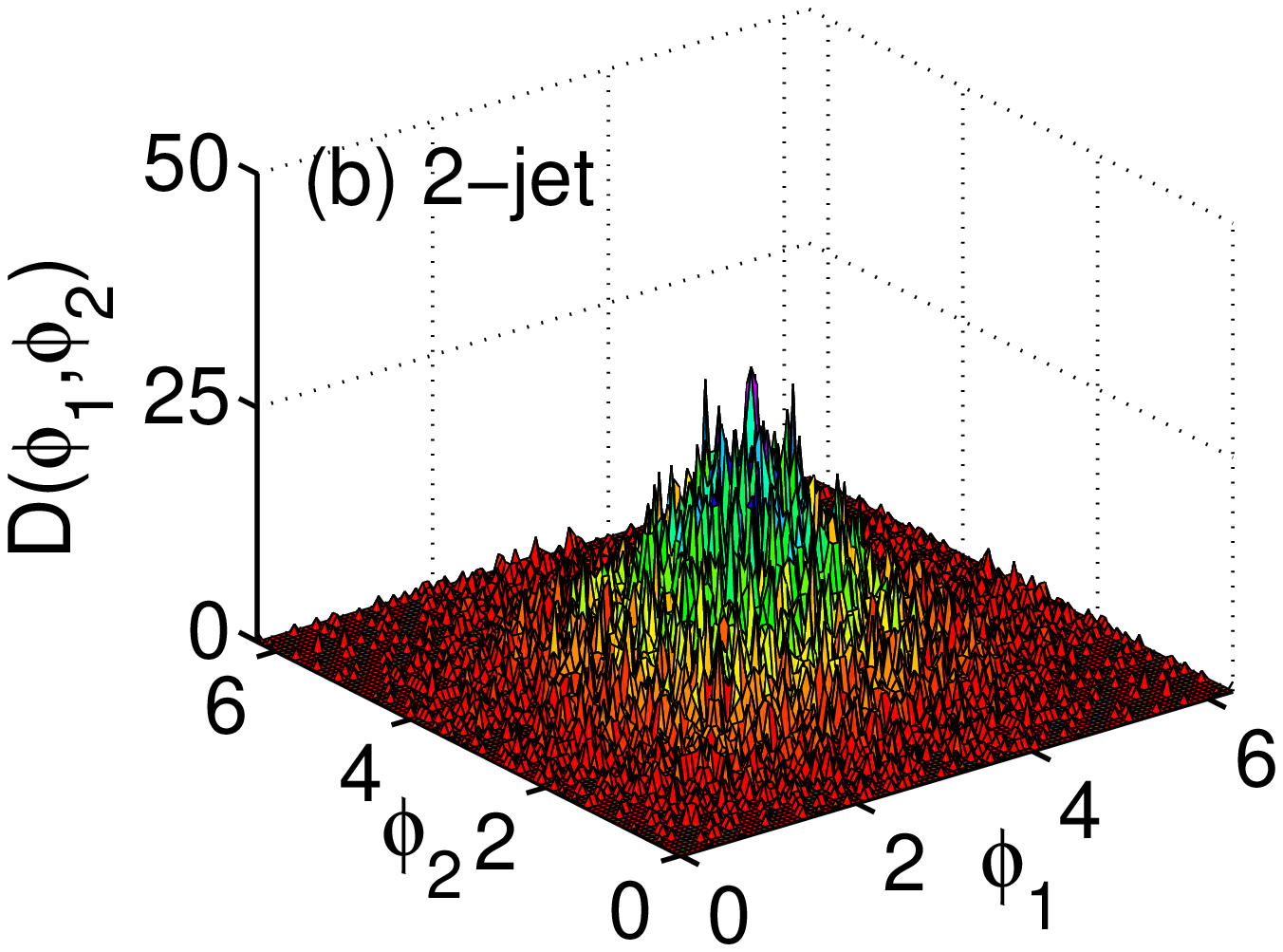}
\caption{Density distributions of hadron pairs for (a) 1-jet and (b) 2-jet triggers.}
\end{figure}

\begin{figure}[tbph]
\vspace*{-6cm}
\includegraphics[width=1\textwidth]{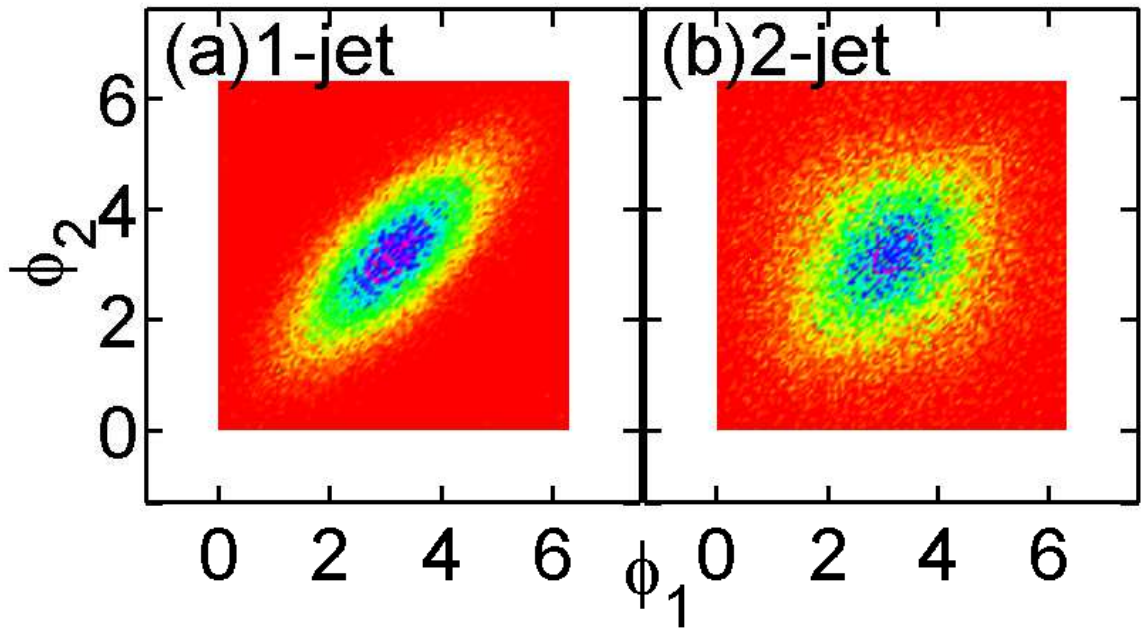}
\vspace*{-1cm}
\caption{Projections of density distributions onto $(\phi_1,\phi_2)$ space for (a) 1-jet and (b) 2-jet triggers.}
\end{figure}

\begin{figure}[tbph]
\vspace*{-2cm}
\includegraphics[width=1\textwidth]{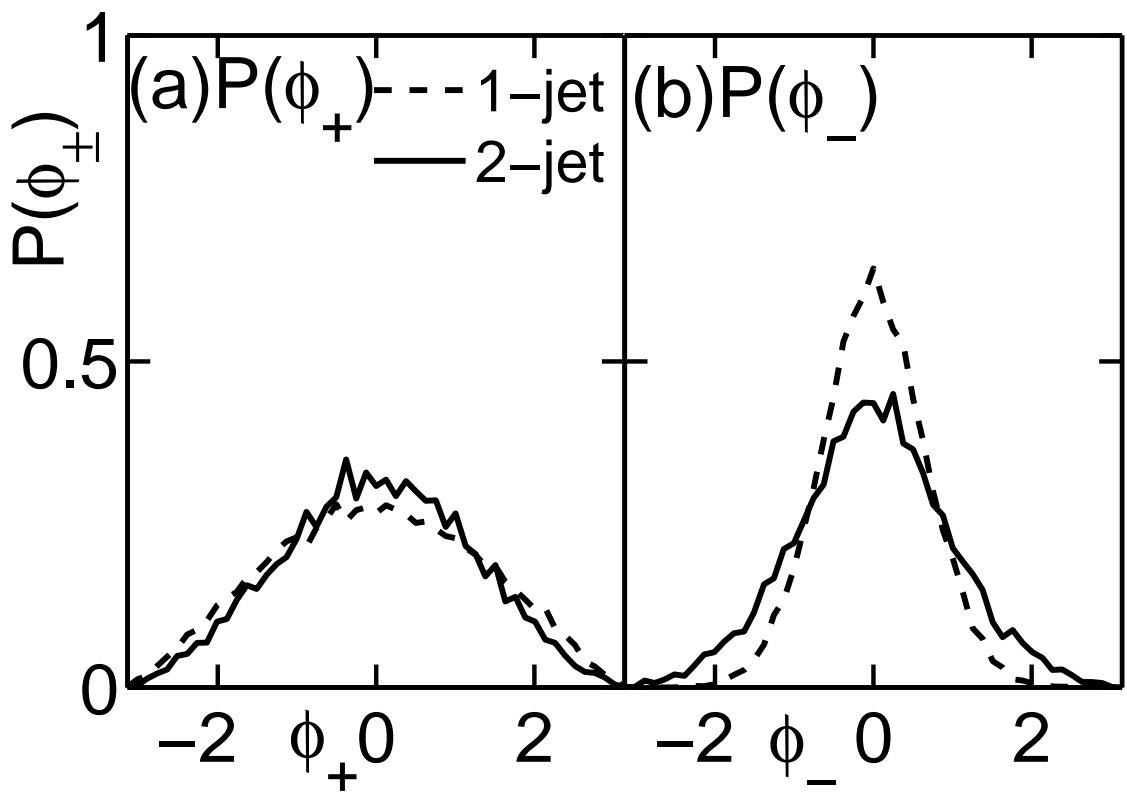}
\vspace*{-1cm}
\caption{Projections of density distributions onto (a) $\phi_+$  and (b) $\phi_-$.}
\end{figure}

\begin{figure}[tbph]
\vspace*{-4.5cm}
\includegraphics[width=1\textwidth,clip]{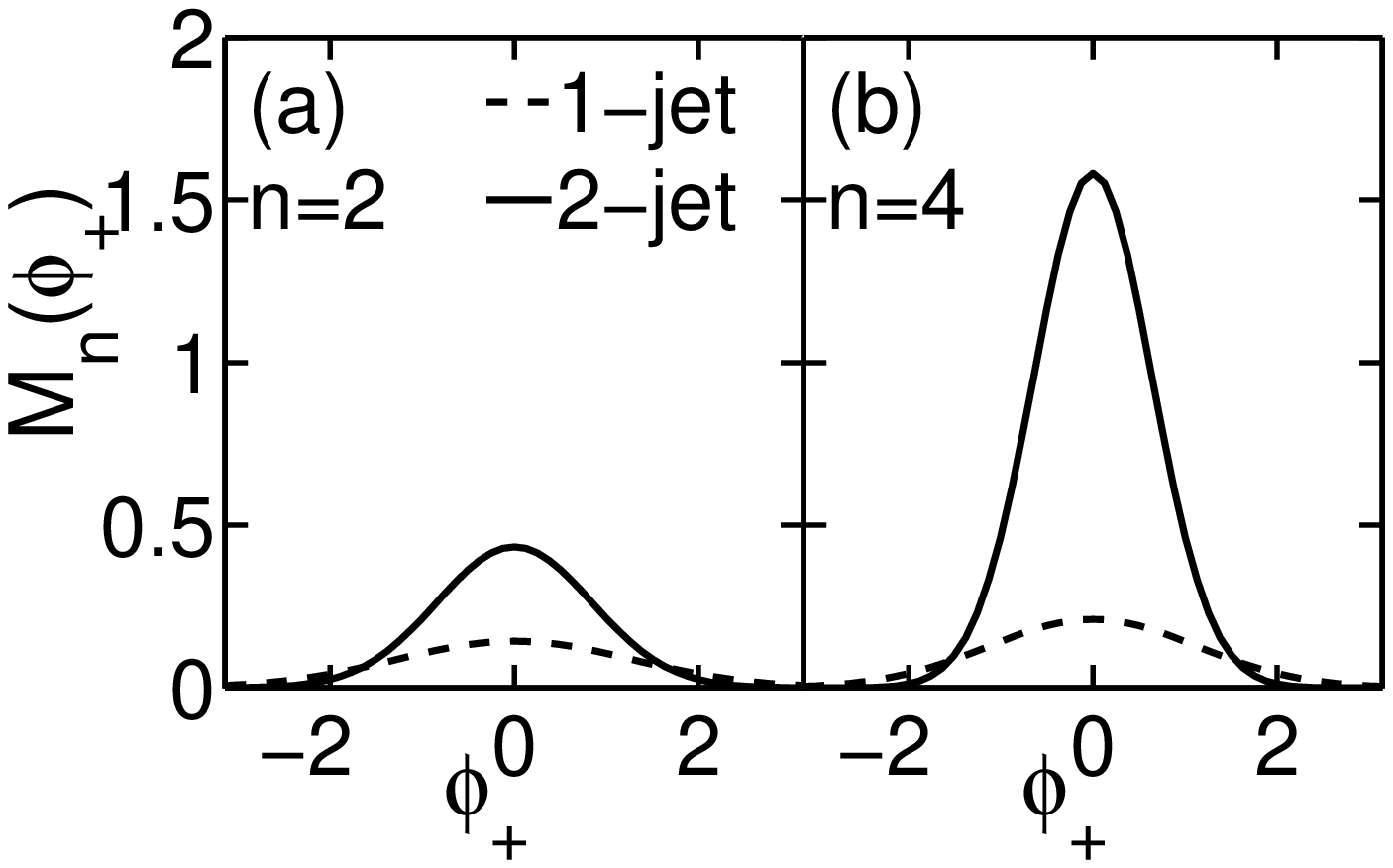}
\vspace*{-1cm}
\caption{Moments $M_n(\phi_+)$ for (a) $n=2$  and (b) $n=4$.}
\end{figure}

\begin{figure}[tbph]
\vspace*{-0.5cm}
\includegraphics[width=1\textwidth,clip]{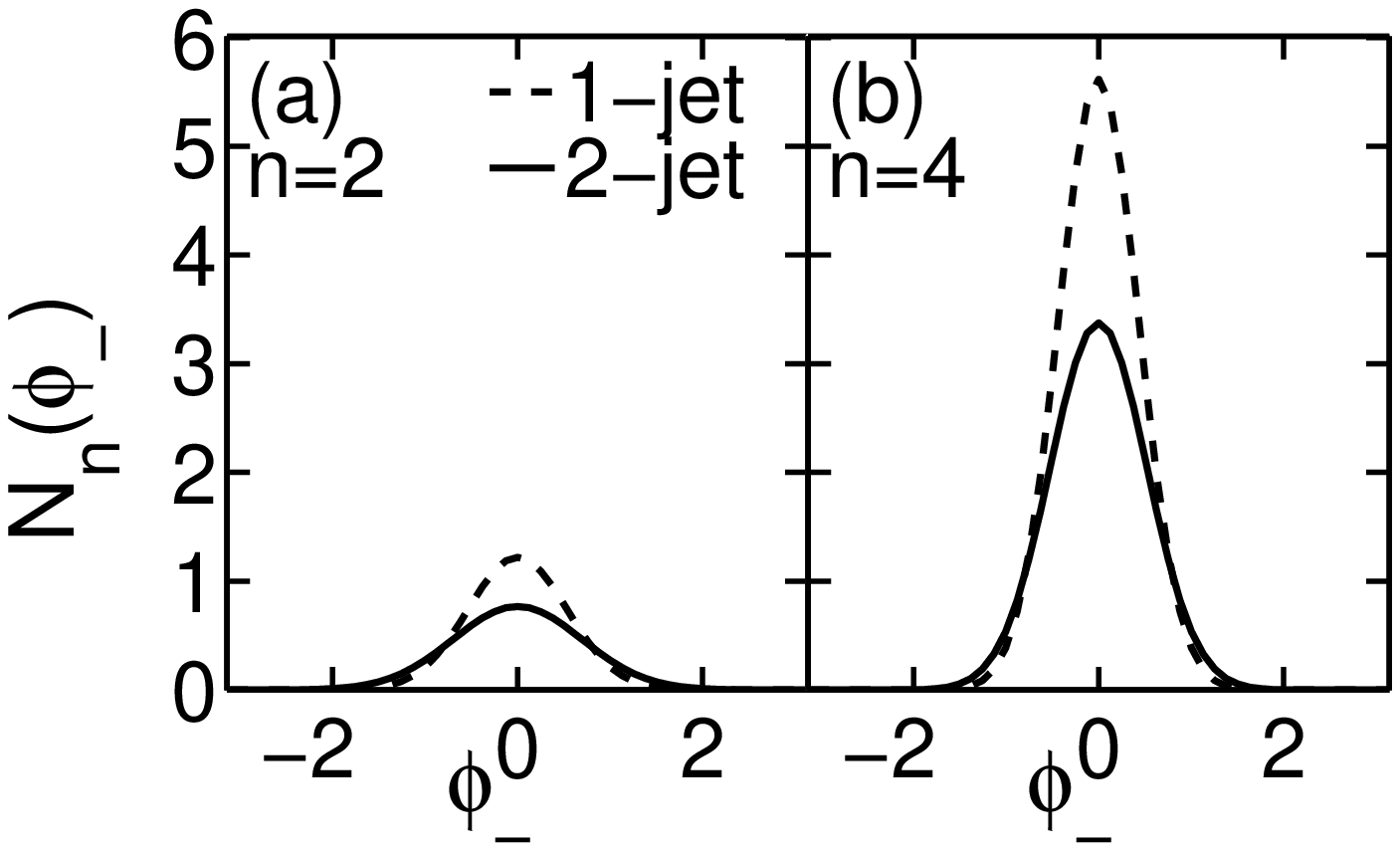}
\vspace*{-2cm}
\caption{Moments $N_n(\phi_-)$ for (a) $n=2$  and (b) $n=4$.}
\end{figure}

\begin{figure}[tbph]
\vspace*{-3cm}
\includegraphics[width=0.8\textwidth]{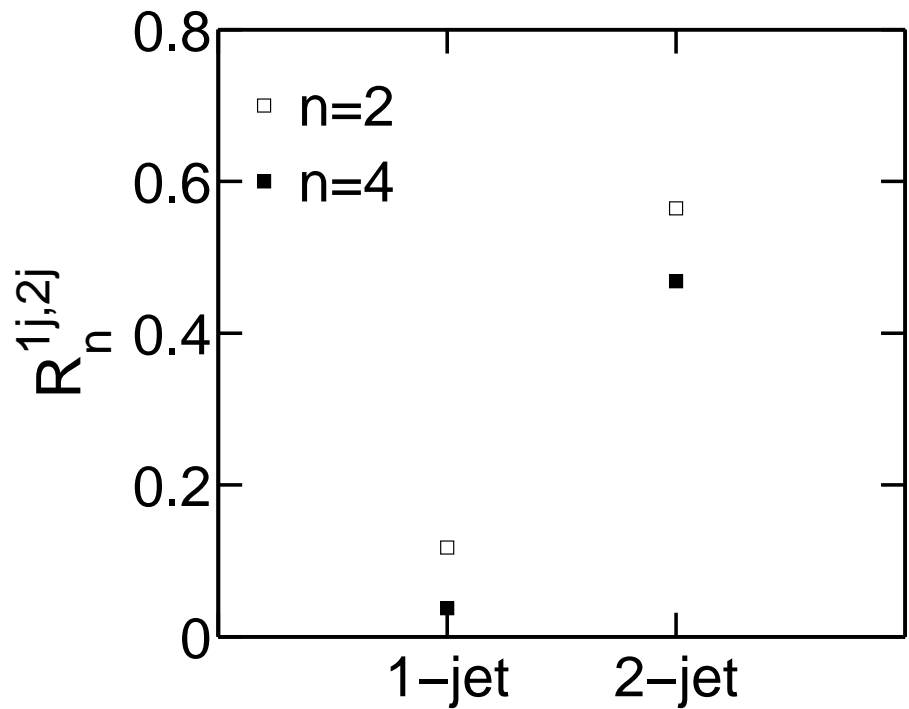}
\caption{Ratio $R_n$ of moments $M_n(0)/N_n(0)$ for 1-jet and 2-jet events for $n=2$  and $n=4$.}
\end{figure}

\begin{figure}[tbph]
\vspace*{-0.5cm}
\includegraphics[width=0.8\textwidth]{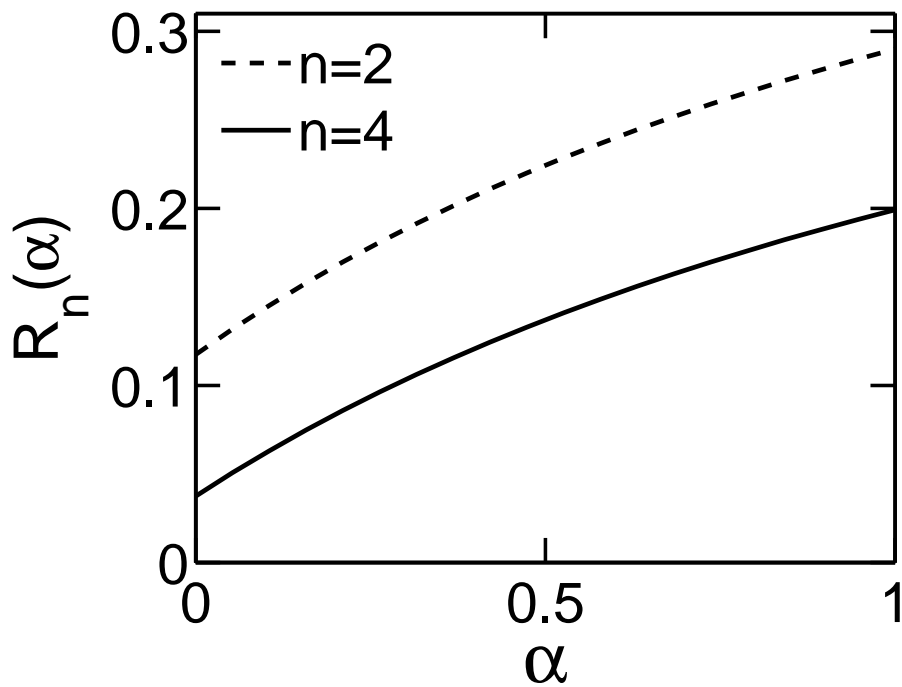}
\vspace*{-.5cm}
\caption{Ratio of moments $R_n(\alpha)$ vs $\alpha$ for $n=2$  and $n=4$.}
\end{figure}

\end{document}